\documentclass{PoS}
\def\ltap{\raisebox{-.4ex}{\rlap{$\,\sim\,$}} \raisebox{.4ex}{$\,<\,$}} 
 
\newcommand\as{\alpha_{\mathrm{S}}}
\def\pthard{p_{\rm T}^{\rm hard}}
\def\ptsoft{p_{\rm T}^{\rm soft}}
\newcommand{\ccaption}[2]{
\begin{center}
\parbox{0.85\textwidth}{
\caption[#1]{\small{{#2}}}
}
\end{center}
}

\title{QCD effects in Higgs boson production at hadron colliders}

\ShortTitle{QCD effects in Higgs boson production at hadron colliders}

\author{\speaker{Massimiliano Grazzini}
\\
        INFN, Sezione di Firenze, I-50019, Sesto Fiorentino, Florence, Italy\\ and\\
        Institute for Theoretical Physics, ETH Zurich, 8093 Zurich, Switzerland\\
        E-mail: \email{grazzini@fi.infn.it}}

\abstract{We present updated predictions for Higgs boson production at the
Tevatron and the LHC and we review their corresponding uncertainties. We
report on
a study of the impact of QCD radiative corrections on the
Higgs boson search at the Tevatron.}

\FullConference{RADCOR 2009 - 9th International Symposium on Radiative Corrections (Applications of Quantum Field Theory to Phenomenology) ,\\
		 October 25 - 30 2009\\
		 Ascona, Switzerland}

\begin{document}

\section{Introduction}

The discovery
of the Standard Model (SM) Higgs boson, or its equivalent in theories beyond the Standard
Model, is one of the main goals of the
experimental program of high-energy colliders.
At the LHC the Higgs boson can be discovered over the full mass range up to $m_H\sim 1$ TeV within a few
years of running. At the Tevatron, the CDF and D0 experiments are now
sensitive to a Higgs signal at $m_H\sim 165$ GeV \cite{tevatron}.

The dominant mechanism for SM Higgs boson production at hadron colliders is gluon-gluon fusion,
through a heavy-quark (mainly, top-quark) loop.
The dynamics of this process
is controlled by strong interactions, and thus
studies of the effect of QCD radiative corrections are necessary to obtain accurate theoretical
predictions.

In QCD perturbation theory the leading order (LO) contribution
to the $gg\to H$ cross section is proportional to $\as^2$,
$\as$ being the QCD coupling.
The QCD corrections have been computed at next-to-leading order (NLO) \cite{Dawson:1991zj,Djouadi:1991tk}
in the heavy-top limit,
and with full dependence on
the masses of the top and bottom quarks \cite{Djouadi:1991tk}.
Next-to-next-to-leading order (NNLO)
corrections have been obtained in the heavy-top limit \cite{NNLOtotal}.
These QCD corrections, which are dominated
by radiation of soft and virtual gluons \cite{soft},
lead to a substantial increase of the LO result.
The QCD computation up to NNLO has been consistently
improved by adding the resummation of soft-gluon logarithmic contributions, up
to next-to-next-to-leading logarithmic (NNLL) accuracy  \cite{Catani:2003zt}.

Recent years have also seen a substantial progress in the computation of
radiative corrections to more exclusive observables
\cite{NNLO3jets,Anastasiou:2004xq,Catani:2007vq,NNLOdy}.
In particular, in the case of Higgs boson
production, two independent fully exclusive NNLO computations are now
available \cite{Anastasiou:2004xq,Catani:2007vq}.

This contribution is divided in two parts. In the first part we discuss an
update for the total cross section \cite{deFlorian:2009hc}, and review the
corresponding uncertainties.
In the second part we consider the fully exclusive NNLO calculation, and
report on a study \cite{Anastasiou:2009bt} of the impact of QCD radiative corrections on the Higgs boson search at the Tevatron.

\section{Total cross section}

In this Section we present an update \cite{deFlorian:2009hc} of the NNLL+NNLO computation of
Ref.~\cite{Catani:2003zt}. 
The results are obtained using the MSTW2008 NNLO partons \cite{Martin:2009iq}.
We first consider the top-quark 
contribution in the loop, and perform the calculation up NNLL+NNLO
in the large-$m_t$ limit. 
The result is rescaled by the exact $m_t$ dependent Born cross section: recent
work has definitely shown that this procedure provides an
excellent approximation (to better than 1\% for $m_H\ltap 300$ GeV)
of the exact top-quark contribution \cite{largemtop}.
We then consider the bottom-quark contribution.
Since in this case the effective theory approach is not applicable,
we follow Ref. \cite{Anastasiou:2008tj} and we include this contribution up to
NLO only \cite{Djouadi:1991tk}.
Finally, we correct the result by including the EW effects \cite{ew} as
evaluated in Ref.~\cite{Actis:2008ug}.
Our central predictions ($\sigma^{\rm best}$) are obtained by setting the
factorization ($\mu_F$) and renormalization ($\mu_R$) scales equal to the Higgs boson mass.
Our results for the Tevatron and the LHC ($\sqrt{s}=14$ TeV) are presented
in Tables 1 and 2, respectively.
Comparing to our previous predictions (see Tables 1 and 2 of Ref.~\cite{Catani:2003zt}), the cross sections change significantly. At the Tevatron the effect
ranges from $+9\%$ for $m_H=115$ GeV to $-9\%$ for $m_H=200$ GeV.
At the LHC  the effect goes from $+30\%$ for $m_H=115$ GeV to $+9\%$ 
for $m_H=300$ GeV, the increase being mainly due to the new MSTW2008 PDFs.


The calculation discussed above is now available through an online calculator
\cite{online},
that can be used to reproduce the results of Tables 1 and 2 or to repeat the
calculation for different Higgs boson masses and/or collider energies.

Our results for the Tevatron can be compared to those presented in 
Ref. \cite{Anastasiou:2008tj}, obtained using the same set of PDFs.
This computation
includes an estimate of mixed QCD-EW contributions. The
main difference with our work arises
in the calculation of the top-quark contribution to the cross section.
In  Ref.~\cite{Anastasiou:2008tj} the latter contribution is computed up to NNLO
but choosing $\mu_F=\mu_R=m_H/2$,
as an attempt to mimic the effects of soft-gluon resummation beyond NNLO.
The final numerical differences at the Tevatron turn out to be small and 
of the order of a few {\em per mille} at the lowest masses, increasing to $2.5\%$ at $m_H=200$ GeV.

The NNLL+NNLO calculation discussed above could be improved in various
respects. Logarithmically enhanced terms beyond NNLL from the Sudakov exponent
have been evaluated in Ref.~\cite{Moch:2005ky,othersoft}. Their
effect, when combined with a full N$^3$LO calculation, can lead to a
reduction of scale uncertainties to about $5\%$ \cite{Moch:2005ky}.
The exact small-$x$ behavior of the NNLO coefficient function
is also known \cite{Marzani:2008az,Harlander:2009my}
and could be included. The numerical effect is, however, smaller than $1\%$
for a light Higgs.
By contrast, the uncertainty that affects the Higgs production cross
section is still large.
The uncertainty basically has two origins:
the one coming from the partonic cross sections, and the one arising from our limited
knowledge of the PDFs.

Uncalculated higher-order QCD radiative corrections are the most important source of uncertainty
on the partonic cross section, and are estimated through scale variations.
The scale uncertainty of our results (see Tables 1 and 2) is about $\pm 9-10\%$ at the Tevatron and ranges from about $\pm 10\%$ ($m_H=110$ GeV)
to about $\pm 7\%$ ($m_H=300$ GeV) at the LHC.
We note that the effect of scale variations
in our resummed calculation
is considerably reduced
with respect to the corresponding NNLO result.
The reduction is more sizeable at the Tevatron, where the resummation effect
is more important.
%

The other important source of uncertainty in the cross section is
the one coming from PDFs.
The MSTW2008 NNLO set provides 40 different grids that allow us
to evaluate the experimental uncertainties.
The outcoming uncertainties (at $90\%$ CL) are reported in Tables 1 and 2.
At the Tevatron the effect ranges from $\pm 6\%$ ($m_H=115$ GeV) to about
$\pm 10\%$ ($m_H=200$ GeV), while at the LHC it
is about $\pm 3\%$
in the mass range we have considered.

A related and important uncertainty is the one coming from the value of the QCD coupling.
Higgs production through gluon fusion starts at ${\cal O}(\as^2)$ and thus
this uncertainty is expected to have a relevant role.
Recently the MSTW collaboration has studied the combined effect of PDF+$\as$
uncertainties \cite{Martin:2009bu}.
We find that at the LHC the PDF+$\as$ uncertainty
is about $7\%$ at 90 \% CL ($m_H\le 300$ GeV),
whereas at the Tevatron it ranges from 7 to 18\% ($m_H\le 200$).
In particular, for $m_H=165$ GeV, we get at the Tevatron $\sigma_{\rm
best}=0.389~{\rm pb}^{+9.2\%}_{-7.7\%}(\rm scale)^{+13.2\%}_{-10.1\%}
(\as+PDF@90\% {\rm CL})$.

{\renewcommand{\arraystretch}{2.0}\renewcommand{\tabcolsep}{1mm}
\begin{table}
\begin{center}
\begin{minipage}{0.31\textwidth}
\begin{tabular}{|c|c|c|c}
\hline
$m_H$ & $\sigma^{\rm best}$ & Scale & PDF\\
\hline
$100$ & $1.861$ & $^{+0.192}_{-0.174}$ & $^{+0.094}_{-0.101}$\\
$105$ & $1.618$ & $^{+0.165}_{-0.149}$ & $^{+0.085}_{-0.091}$\\
$110$ & $1.413$ & $^{+0.142}_{-0.127}$ & $^{+0.077}_{-0.083}$\\
$115$ & $1.240$ & $^{+0.123}_{-0.110}$ & $^{+0.070}_{-0.075}$\\
$120$ & $1.093$ & $^{+0.107}_{-0.095}$ & $^{+0.065}_{-0.069}$\\
$125$ & $0.967$ & $^{+0.094}_{-0.083}$ & $^{+0.059}_{-0.063}$\\
$130$ & $0.858$ & $^{+0.082}_{-0.072}$ & $^{+0.054}_{-0.058}$\\
\hline
\end{tabular}
\end{minipage}
\begin{minipage}{0.31\textwidth}
\begin{tabular}{|c|c|c|c}
\hline
$m_H$ & $\sigma^{\rm best}$ & Scale & PDF\\
\hline
$135$ & $0.764$ & $^{+0.073}_{-0.063}$ & $^{+0.050}_{-0.053}$\\
$140$ & $0.682$ & $^{+0.065}_{-0.056}$ & $^{+0.046}_{-0.049}$\\
$145$ & $0.611$ & $^{+0.057}_{-0.049}$ & $^{+0.042}_{-0.045}$\\
$150$ & $0.548$ & $^{+0.051}_{-0.044}$ & $^{+0.039}_{-0.042}$\\
$155$ & $0.492$ & $^{+0.045}_{-0.039}$ & $^{+0.036}_{-0.038}$\\
$160$ & $0.439$ & $^{+0.040}_{-0.034}$ & $^{+0.033}_{-0.035}$\\
$165$ & $0.389$ & $^{+0.036}_{-0.030}$ & $^{+0.030}_{-0.032}$\\
\hline
\end{tabular}
\end{minipage}
\begin{minipage}{0.31\textwidth}
\begin{tabular}{|c|c|c|c}
\hline
$m_H$ & $\sigma^{\rm best}$ & Scale & PDF\\
\hline
$170$ & $0.349$ & $^{+0.032}_{-0.027}$ & $^{+0.028}_{-0.029}$\\
$175$ & $0.314$ & $^{+0.029}_{-0.024}$ & $^{+0.026}_{-0.027}$\\
$180$ & $0.283$ & $^{+0.026}_{-0.021}$ & $^{+0.024}_{-0.025}$\\
$185$ & $0.255$ & $^{+0.023}_{-0.019}$ & $^{+0.022}_{-0.023}$\\
$190$ & $0.231$ & $^{+0.021}_{-0.017}$ & $^{+0.020}_{-0.021}$\\
$195$ & $0.210$ & $^{+0.019}_{-0.015}$ & $^{+0.019}_{-0.020}$\\
$200$ & $0.192$ & $^{+0.017}_{-0.014}$ & $^{+0.018}_{-0.019}$\\
\hline
\end{tabular}
\end{minipage}
\ccaption{}{\label{tab:tev}{\em Cross sections (in pb)
at the Tevatron ($\mu_F=\mu_R=m_H$) with $\sqrt{s}=1.96$ TeV 
using the MSTW2008 \cite{Martin:2009iq} parton densities.}}
\end{center}
\end{table}}

{\renewcommand{\arraystretch}{2.0}\renewcommand{\tabcolsep}{1mm}
\begin{table}
\begin{center}
\begin{minipage}{0.31\textwidth}
\begin{tabular}{|c|c|c|c}
\hline
$m_H$ & $\sigma^{\rm best}$ & Scale & PDF\\
\hline
$100$ & $ 74.58 $& $^{+7.18}_{-7.54}$ & $^{+1.86}_{-2.45}$\\
$110$ & $ 63.29 $& $^{+5.87}_{-6.20}$ & $^{+1.54}_{-2.02}$\\
$120$ & $ 54.48 $& $^{+4.88}_{-5.18}$ & $^{+1.30}_{-1.70}$\\
$130$ & $ 47.44 $& $^{+4.12}_{-4.38}$ & $^{+1.12}_{-1.45}$\\
$140$ & $ 41.70 $& $^{+3.47}_{-3.75}$ & $^{+0.97}_{-1.25}$\\
$150$ & $ 36.95 $& $^{+3.02}_{-3.24}$ & $^{+0.85}_{-1.10}$\\
$160$ & $ 32.59 $& $^{+2.60}_{-2.79}$ & $^{+0.73}_{-0.97}$\\
\hline
\end{tabular}
\end{minipage}
\begin{minipage}{0.31\textwidth}
\begin{tabular}{|c|c|c|c}
\hline
$m_H$ & $\sigma^{\rm best}$ & Scale & PDF\\
\hline
$170$ & $ 28.46 $& $^{+2.22}_{-2.39}$ & $^{+0.65}_{-0.84}$\\
$180$ & $ 25.32 $& $^{+1.92}_{-2.08}$ & $^{+0.58}_{-0.74}$\\
$190$ & $ 22.63 $& $^{+1.68}_{-1.83}$ & $^{+0.52}_{-0.66}$\\
$200$ & $ 20.52 $& $^{+1.49}_{-1.63}$ & $^{+0.48}_{-0.60}$\\
$210$ & $ 18.82 $& $^{+1.34}_{-1.47}$ & $^{+0.45}_{-0.55}$\\
$220$ & $ 17.38 $& $^{+1.22}_{-1.33}$ & $^{+0.42}_{-0.51}$\\
$230$ & $ 16.15 $& $^{+1.11}_{-1.22}$ & $^{+0.39}_{-0.48}$\\
\hline
\end{tabular}
\end{minipage}
\begin{minipage}{0.31\textwidth}
\begin{tabular}{|c|c|c|c}
\hline
$m_H$ & $\sigma^{\rm best}$ & Scale & PDF\\
\hline
$240$ & $ 15.10 $& $^{+1.03}_{-1.12}$ & $^{+0.37}_{-0.45}$\\
$250$ & $ 14.19 $& $^{+0.95}_{-1.04}$ & $^{+0.36}_{-0.43}$\\
$260$ & $ 13.41 $& $^{+0.88}_{-0.97}$ & $^{+0.35}_{-0.41}$\\
$270$ & $ 12.74 $& $^{+0.83}_{-0.91}$ & $^{+0.33}_{-0.39}$\\
$280$ & $ 12.17 $& $^{+0.78}_{-0.86}$ & $^{+0.33}_{-0.38}$\\
$290$ & $ 11.71 $& $^{+0.74}_{-0.82}$ & $^{+0.32}_{-0.37}$\\
$300$ & $ 11.34 $& $^{+0.71}_{-0.78}$ & $^{+0.32}_{-0.36}$\\
\hline
\end{tabular}
\end{minipage}
\ccaption{}{\label{tab:lhc14}{\em Cross sections (in pb)
at the LHC ($\mu_F=\mu_R=m_H$) with $\sqrt{s}=14$ TeV 
using the MSTW2008 \cite{Martin:2009iq} parton densities.}}
\end{center}
\end{table}}

We finally point out that, besides MSTW, we have at present only two other
NNLO parton analyses: ABKM09 \cite{Alekhin:2009ni} and JR09VFNNLO
\cite{JimenezDelgado:2009tv}.
A comparison of the central values of the cross section shows that at the LHC
ABKM09 (JR09VFNNLO) result is smaller than MSTW by about 7\% (11\%) for
$m_H=115$ GeV and by 11\% (8\%) for $m_H=300$ GeV.
At the Tevatron ABKM09 (JR09VFNNLO) result is smaller than MSTW by about 26\%
(2\%) for $m_H=165$ GeV. Although these three NNLO sets are obtained through
different approaches, the large differences in the corresponding results confirm that
the uncertainty in the total Higgs production cross section is still large
and, at least at the Tevatron, dominated by PDFs.

\section{QCD effects in $H\to WW\to l\nu l\nu$ at the Tevatron}

In the previous Section we have discussed in detail perturbative predictions
for the fully inclusive Higgs production cross section. Total cross sections,
however, are ideal quantities: experiments have always a finite acceptance.
To properly take into account the kinematical cuts applied in the experimental
analysis, {\em fully exclusive} calculations are actually needed.
LO calculations are in this respect straightforward: one can compute the relevant matrix element and
integrate it numerically over the multiparton phase space.
Beyond LO the QCD computation is affected by infrared singularities that prevent a
straightforward implementation of numerical techniques.

In particular, at NNLO, only few fully exclusive computations exist, due to
their substantial conceptual and
technical complications \cite{NNLO3jets,Anastasiou:2004xq,Catani:2007vq,NNLOdy}.
For Higgs boson production through gluon fusion, two fully
independent
computations exist that are
implemented in available numerical codes \cite{Anastasiou:2004xq,Catani:2007vq}.

The CDF and D0 collaborations at the Tevatron have recently reported a
combination of their results up to 5.4 fb$^{-1}$. According to this combination,
a SM Higgs boson of masses between 163 and 166 GeV is excluded at 95\% CL.
In this mass region the signal is dominated by the $H\to WW\to l\nu l\nu$
channel, which provides a small number of events over a huge background.
After a first cut-based selection, background
processes remain dominant and processing of real data and Monte-Carlo simulations with
Artificial Neural Network (ANN) methods follows.
Given the sensitivity of the gluon fusion cross-section to higher order effects, it is important
to establish that the sophisticated methods used in the Tevatron analysis account
for these effects within the estimated uncertainties.

In Ref.\cite{Anastasiou:2009bt} we have performed a study of the inpact of cuts on the Higgs
boson signal. We assume a Higgs boson mass $m_H=160$ GeV and
apply the following cuts. We consider the
$H\to WW\to \mu^+\mu^-\nu{\bar \nu}$ channel and require at least one lepton with
$p_T>20$ GeV and $|\eta|<1.1$. The invariant mass of the charged leptons
should be $m_{ll}>16$ GeV. Leptons should be isolated: the total transverse
energy in a cone of radius R=0.4 should be smaller than $10\%$ of the lepton
$p_T$. Jets are defined according to the $k_T$ algorithm with $D=0.4$: a jet
is required to have $p_T>15$ GeV and $|\eta|<3$.

We define the variable $\mathrm{MET}^*$ as
\begin{equation}
\mathrm{MET}^*=\left\{\begin{array}{l} \mathrm{MET}\;\;\;\;\;\;\;\;\;\;\;\;,\;\phi\geq\pi/2  \\ \mathrm{MET}\times\sin{\phi}\;,\;\phi<\pi/2 \end{array}\right.,
\end{equation}
where $\phi$ is the angle in the transverse plane between the missing
transverse energy, MET, and the nearest charged lepton or jet.
We require $\mathrm{MET}^*>25$ GeV, which suppresses the background from Drell-Yan lepton pairs
and removes contributions from mismeasured leptons or jets.
To suppress the $t{\bar t}$ background, we require at most one hadronic jet.

With the cuts discussed above the NLO K-factor is reduced from 2.41 to 2.15
and the NNLO K-factor is reduced from 3.31 to 2.59 (for $\mu_F=\mu_R=m_H$).
In order to study the stability of perturbative corrections in the presence of
these cuts we have studied a set of kinematical distributions that can be computed with
our NNLO programs. We consider the transverse momenta of the leading and
trailing lepton $\pthard$, $\ptsoft$, the invariant mass of the charged
lepton, $m_{ll}$, the azimuthal separation of the charged leptons in the
transverse plane and the missing transverse energy, MET.

A study of these distributions up to NNLO does not show significant
instabilities. We have then compared the results to the same distributions
obtained with PYTHIA and MC@NLO, rescaled so as to match the total NNLO cross
section, without finding significant differences.

An essential part of the experimental studies concerns
distributions of discrimination variables, defined via ANNs.
In Ref.~\cite{Anastasiou:2009bt} we have studied for
the first time an ANN output distribution up to NNLO in perturbation theory,
using as input variables the leptonic quantities defined above.
The results confirm the agreement
discussed above.

Despite the agreement in the shape of the leptonic distributions, confirmed
by the ANN analysis, a comparison of the acceptances show some discrepancy.
The acceptance obtained with HERWIG and MC@NLO is consistent with the
acceptance from the NNLO calculation.
In contrast, we find
that the acceptance computed with PYTHIA is between 12\% and 21\% smaller than
the NNLO acceptance, depending on the choice of the renormalization and
factorization scales. This result is not significantly altered by
hadronization and underlying event effects and appears
instead to be related to the matrix element and parton shower implementation
in PYTHIA itself.

%
%

\section{Summary}

We have presented updated predictions for Higgs boson production at
the Tevatron and the LHC, and discussed their uncertainties.
We have presented a study of the impact of QCD radiative corrections on the
Higgs search in the $H\to WW\to l\nu l\nu$ channel
at the Tevatron, based on the NNLO
calculations of Refs.~\cite{Anastasiou:2004xq,Catani:2007vq}.
This study shows that these NNLO
programs can provide an essential help in the validation of the results from
standard Monte Carlo event generators and in the assessment of theoretical
uncertainties.

\noindent{\bf Acknowledgements}

I would like to thank the organizers of RADCOR 2009 for their
kind invitation and for financial support.

\end{document}